\documentclass[12pt]{iopart}
\usepackage{graphicx}
\usepackage{iopams}
\begin{document}

\title{Properties of Quarkonia at $T_{\rm c}$}

\author{Su Houng Lee and Kenji Morita}

\address{Institute of Physics and Applied Physics,
Yonsei University, Seoul 120-749, Korea}
\ead{suhoung@yonsei.ac.kr}
\begin{abstract}
We discuss how the spectral changes of quarkonia  at $T_{\rm c}$ can reflect
 the ``critical'' behaviour of QCD phase transition.
Starting from the temperature dependencies of the energy density and
pressure from lattice QCD calculation, we extract the temperature
dependencies of the scalar and spin 2 gluon condensates near $T_{\rm c}$.  We
 also parameterize these changes into the electric and magnetic
 condensate near $T_{\rm c}$.  While the magnetic condensate hardly changes
 across $T_{\rm c}$, we find that the electric condensate increases abruptly
 above $T_{\rm c}$.   Similar abrupt change is also seen in the scalar
 condensate.  Using the QCD second-order Stark effect and QCD sum rules,
 we show that these sudden changes induce equally abrupt changes in the
 mass and width of $J/\psi$, both of which are larger than 100 MeV at
 slightly above $T_{\rm c}$.

\end{abstract}

%Uncomment for PACS numbers title message
%\pacs{12.38.-t, 12.38.Mh, 25.75.Nq}
% Keywords required only for MST, PB, PMB, PM, JOA, JOB?
%\vspace{2pc}
%\noindent{\it Keywords}: Article preparation, IOP journals
% Uncomment for Submitted to journal title message
%\submitto{\JPA}
% Comment out if separate title page not required
%\maketitle

\section{Introduction}

$J/\psi$ emanating from heavy ion collision is considered to be an
intriguing tool to investigate the properties of quark gluon plasma
(QGP) expected to form  in the early stages of a heavy ion
collision~\cite{hashimoto86,Matsui_PLB178}.   Indeed measurements at past SPS
and recent RHIC data show nontrivial suppression patterns that could be
consistent with the original prediction that the $J/\psi$ will dissolve
in QGP.  However, the subject has gained a new turn recently as lattice
calculations show that $J/\psi$ will survive past the critical
temperature $T_{\rm c}$ up to about 1.6$T_{\rm c}$, while $\chi_c$ and
$\psi'$ will dissolve just above $T_{\rm c}$~\cite{Asakawa_PRL92,Datta_PRD69}.
The notion that $J/\psi$
will not dissolve immediately above $T_{\rm c}$ was also considered
before~\cite{hansson88}, as it was known that non-perturbative aspects of
QCD persist in QGP~\cite{manousakis87,Lee89}.  These results suggest that
the sudden disappearance of $J/\psi$ at $T_{\rm c}$ is not the order parameter
of QCD phase transition.  On the other hand, the phase transition is
accompanied by sudden changes in the chiral condensate, the heavy quark
potential and the energy density, and should therefore have some effect
on the properties of quarkonia at $T_{\rm c}$.

Unfortunately, the lattice results based on the maximum entropy method have
large error bars, and one can not determine the detailed properties of
quarkonium above $T_{\rm c}$, except its existence.   Solving the
Schr\"{o}dinger equation with potential extracted from the lattice is one
possibility~\cite{Wong05}.
However, here there are some controversies on how to extract the correct
potential from the free energy calculations. Even with a chosen
prescription, one has to solve the Schr\"{o}dinger equation at discrete
temperatures steps, and it becomes not clear how the sudden critical
behaviour is translated into the discrete temperature steps.  Thermal
perturbation alone is also not sufficient to probe the temperature
region from $T_{\rm c}$ to 2.5$T_{\rm c}$\cite{Blaizot03}, which is now known to be strongly interacting.

Therefore, a systematic non-perturbative method is essential to treat the phase transition region.  Here, we will summarize the recent
developments~\cite{morita_jpsi,lee_morita_stark}
to attach the problem using QCD sum rules and the QCD second order Stark
effect. The inputs are the temperature dependencies of local gluonic
operators, which undergoes abrupt changes across the phase transition as
does the energy density.

\section{The gluon matter}

The order parameter for QCD phase transition is the thermal Wilson line
for pure gauge theory, and the quark condensates for QCD with
massless quarks.  For realistic QCD none of them are order parameters in
the strict sense.  But simulation shows that the sudden changes in both
parameters take places at the same critical temperature $T_{\rm c}$,
which is determined by the susceptibilities.  At the same $T_{\rm c}$,
the energy density also makes a drastic change, which is somehow
universal for any quark flavours. The sudden changes apparent in the
energy density and pressure can be translated to temperature
dependencies of local gluonic operators, which are expected to embody
the non perturbative nature of QCD.  The link is obtained through the
energy momentum tensor, which can be written in terms of symmetric
traceless part (twist-2 gluon) and the trace part (gluon condensate).

\begin{eqnarray}
T^{\alpha\beta}  = -{\cal S T}( G^{a\alpha\lambda}G^{a\beta}_{\lambda} )
+\frac{g^{\alpha \beta}}{4} \frac{\beta(g)}{2g}G^{a}_{\mu\nu}G^{a\mu\nu} ,
\label{em-tensor}
\end{eqnarray}
where we will use the LO beta function
$\beta(g)=-\frac{g^3}{(4 \pi)^2}(11-\frac{2}{3}N_{\rm f})$.
The thermal expectation value of this operator can be related to the
lattice measurements of energy-momentum tensor through the following
relation.
\begin{equation}
\left\langle T^{\alpha \beta} \right\rangle_T
 = (\varepsilon+p)\left(u^\alpha u^\beta - \frac{1}{4}g^{\alpha \beta}\right)
 + \frac{1}{4}(\varepsilon-3p)g^{\alpha\beta}.\label{eq:e-m-t}
\end{equation}
Here, $u^\alpha$ is the four velocity of the heat bath.  With the
following definitions,
\begin{eqnarray}
\left\langle \frac{\alpha_{\rm s}}{\pi}G^{a}_{ \mu \nu}
 G^{a \mu \nu }\right\rangle_T & = & G_0(T)
 \label{eq:defG0}\\
\left\langle {\cal S T}(  \frac{\alpha_{\rm s}}{\pi} G^{a\alpha\lambda}G^{a\beta}_{\lambda} )\right\rangle_T
  &= &
\left(u^\alpha u^\beta -
				    \frac{1}{4}g^{\alpha \beta}\right)  G_2(T)   ,\label{eq:defg2}
\end{eqnarray}
we find for pure SU(3) gauge theory,
\begin{eqnarray}
G_0(T)  =   G_0^{\rm vac}-\frac{8}{11}(\varepsilon-3p), \quad
G_2(T)  =   -\frac{\alpha_{\rm s}(T)}{\pi}(\varepsilon+p),
\end{eqnarray}
where $G_0^{\rm vac}$ is the value of the scalar gluon condensate in
vacuum.

At low temperature, the scalar part is dominated and characterized by the non-perturbative
contribution of the QCD vacuum.
%Here, even at higher temperature, the ideal gas contribution vanishes.
At high temperature, its behaviour will be dominated by the quasi-particle
contribution.  Just above $T_c$, it is
still dominated by the non-perturbative contribution.
For the heat bath at rest, one can rewrite the thermal expectation values in terms of electric and magnetic condensate.
\begin{eqnarray}
\left\langle \frac{\alpha_{\rm s}}{\pi} \boldsymbol{E}^2
\right\rangle_T & = & -\frac{1}{4}G_0(T) -\frac{3}{4} G_2(T)
 \label{eq:defe2}\\
\left\langle \frac{\alpha_{\rm s}}{\pi} \boldsymbol{B}^2 \right\rangle_T
  &= & \frac{1}{4}G_0(T) -\frac{3}{4} G_2(T) .\label{eq:defb2}
\end{eqnarray}
Figure 1 shows the temperature dependence of $G_0$, $G_2$ or
$\boldsymbol{E}^2$ and $\boldsymbol{B}^2$.   One should first note that
the sudden increase of energy density is translated to the anomalously
large and sudden decrease in $G_0$, which deviates largely from the
asymptotic $T^4$ behaviour.  In contrast, $G_2$ reaches the asymptotic
value quickly.  Similarly, one finds that there is a sudden increase in
the electric condensate $\boldsymbol{E}^2$, while the magnetic condensate
$\boldsymbol{B}^2$ hardly changes above $T_c$.   This can be related to
the fact that the area law behaviour of the space time Wilson loop
changes to the perimeter law above $T_c$, while that of the space-space
Wilson loop does not~\cite{manousakis87}.  The connection comes in through
the operator product expansion (OPE) of rectangular Wilson loop, which was found to be expressible in terms of the electric condensates and the  magnetic condensates for the space-time and space-space Wilson loops respectively~\cite{shifman80}.  To investigate the consequences of the abrupt changes of condensates to the properties of $J/\psi$, we will use perturbative QCD and QCD sum rules.

\begin{figure}
 \begin{center}
  \includegraphics[width=0.4\textwidth]{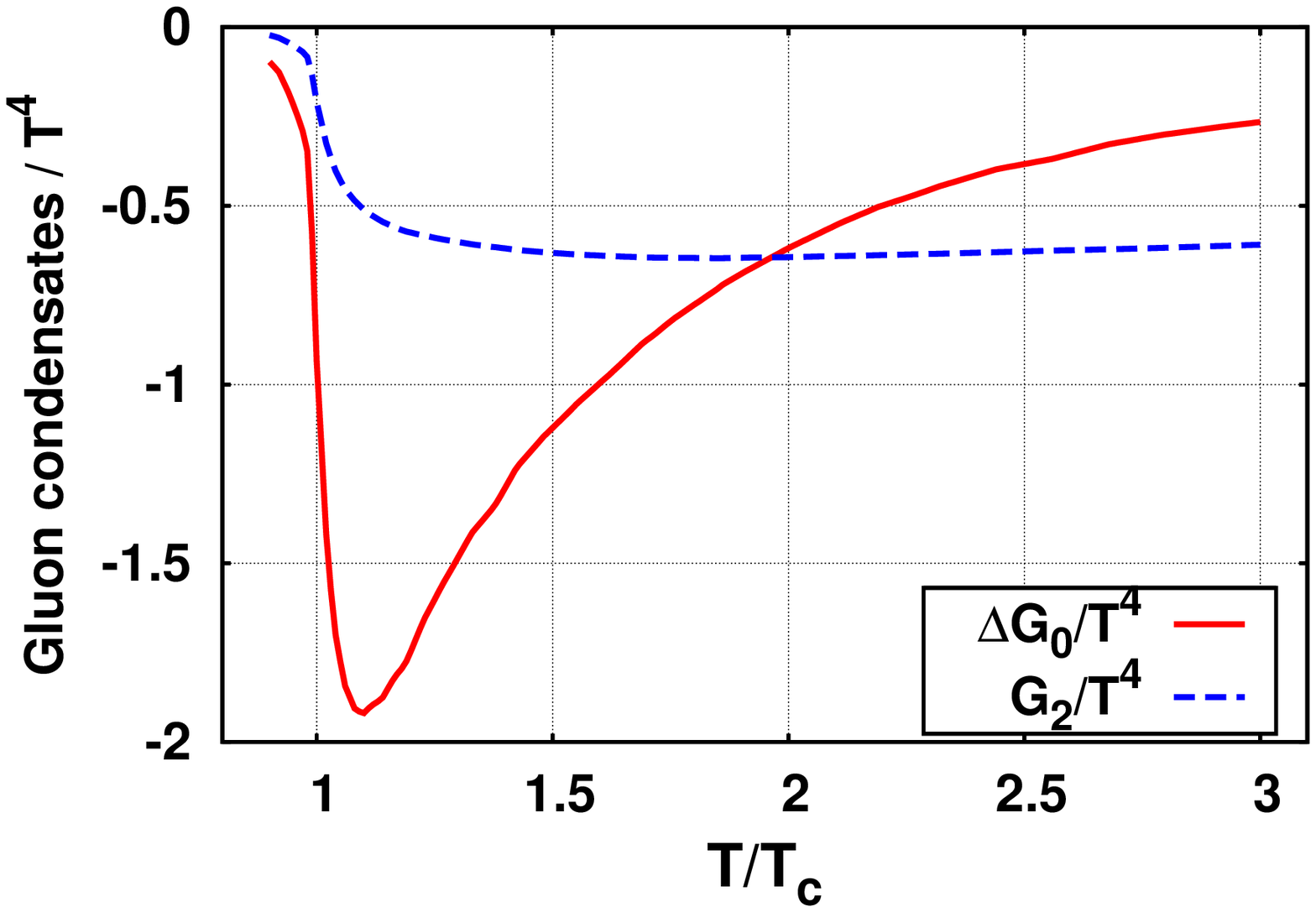}\includegraphics[width=0.4\textwidth]{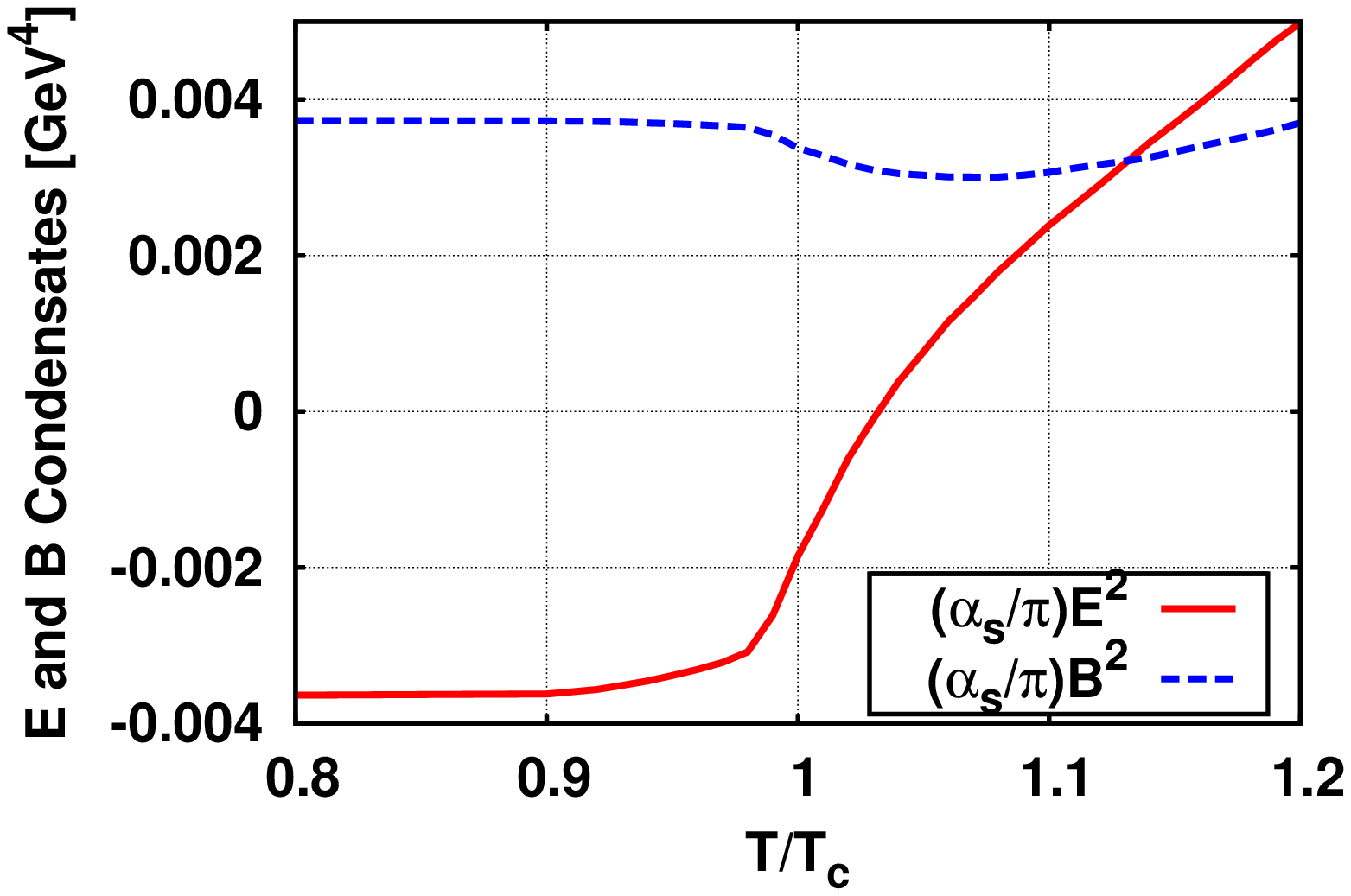}
  \caption{ a) Gluon condensates and b) electric and
  magnetic condensates  near $T_{{\rm c}}$.}
  \label{fig:gc}
 \end{center}
\end{figure}

\section{Results from perturbative QCD}

The perturbative QCD formalism for calculating the interaction between
heavy quarkonium and partons was first developed by
Peskin~\cite{peskin79,bhanot79}.    The counting scheme in this formalism is
obtained in the non-relativistic limit.   Because of this, the formula
for the mass shift reduces to the second-order Stark effect in QCD,
which was used to calculate the mass shift of charmonium in nuclear
matter~\cite{luke92}. The information needed from the medium is
the electric field square.  Noting that the dominant change across the
phase transition is that of the electric condensate, one finds that the
QCD second-order Stark effect is the most natural formula to be used across
the phase transition.

The QCD second-order Stark effect for the ground state charmonium with
momentum space wave function normalized as
$\int \frac{d^3p}{(2 \pi)^3} |\psi(\bf{p})|^2=1$ is as follows,
\begin{eqnarray}
\delta m_{J/\psi} & = &
 - \frac{1}{18} \int_0^\infty dk^2
 \left| \frac{\partial \psi(k)}{\partial k}
 \right|^2
 {k \over k^2/m_c+ \epsilon}
 \left\langle \frac{\alpha_{\rm s}}{\pi} \Delta\boldsymbol{E}^2
	\right\rangle_T \nonumber \\
& = & -\frac{ 7\pi^2}{18} \frac{a^2}{\epsilon}
 \left\langle \frac{\alpha_{\rm s}}{\pi} \Delta\boldsymbol{E}^2
 \right\rangle_T , \label{stark}
\end{eqnarray}
where $k=|\boldsymbol{k}|$ and
$\langle \frac{\alpha_{\rm s}}{\pi}\Delta \boldsymbol{E}^2 \rangle_T$
denotes the value of change of the electric condensate from its vacuum
value so that $\delta m_{J/\psi} = 0$ in vacuum.
The second line is obtained for the Coulomb wave function.
Here, $\epsilon$ is the  binding energy and $m_c$ the charm quark mass.
These parameters are fit to the size of the wave function obtained in
the potential model~\cite{eichten80}, and to the mass of $J/\psi$
assuming it to be a Coulombic bound state in the heavy quark
limit~\cite{peskin79}.  The fit
gives $m_c=1704$ MeV, $a=0.271$ fm and $\alpha_{\rm s}=0.57$.    Few comments
are in order.  The minus sign in \eref{stark} is a model
independent result and follows from the fact that the second-order Stark
effect is negative for the ground state.  Second, the important parts of
the formula are the Bohr radius $a$, the binding energy $\epsilon$ and
the temperature dependence of the electric condensate squared.  In the
formula, the factor of $a^2$ follows from the dipole nature of the
interaction, and the binding energy from the inverse propagator,
characterizing the separation scale~\cite{peskin79,oh02}.  Therefore,
the actual value of the mass shift does not depend much on the form of
the wave function as long as the size of the wave function is fixed.

%%%%%%%%%%%%%%%%%%%%%%%%%%%%%%%%%%%%%%%
\begin{figure}[tb]
\begin{center}
\includegraphics[width=0.5\textwidth]{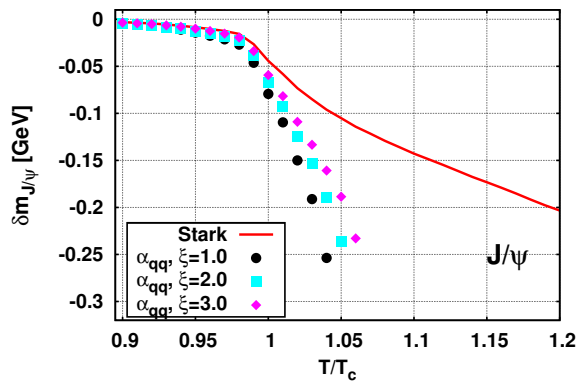}
\end{center}
\caption{Mass shift from the second-order Stark effect (solid line) and
 the maximal mass shift obtained from QCD sum rules from
 Ref.~\cite{morita_jpsi} (points).
}
\label{mass-shift}
\end{figure}
%%%%%%%%%%%%%%%%%%%%%%%%%%%%%%%%%%%%%%%

The solid line in Fig.~\ref{mass-shift} shows the mass shift obtained
from the second-order Stark effect with the extracted lattice input for the electric condensate shown in Fig. 1-b.  As the formula is based on
multipole expansion, it will break down if the higher dimensional
condensates become large.  At this moment, there is no lattice
calculation for the temperature dependence of higher dimensional gluonic
operators.  However, as discussed before, up to 1.1$T_{\rm c}$ the temperature
dependence is dominated by the sudden increase in the electric condensate
with little change in the magnetic counterpart.  Moreover, the OPE in
the QCD sum rules were found to be reliable up to
1.05$T_{\rm c}$~\cite{morita_jpsi}.  Therefore
our result should at
least be valid up to the same temperature.   Above that temperature, the change of the electric condensate amounts to more than 100\% of its vacuum value and higher order corrections should be important. As can be seen in
Fig. 1, the mass reduces abruptly above $T_{{\rm c}}$ and becomes smaller by
about 100 MeV at 1.05$T_{\rm c}$, reflecting the critical behaviour of
the QCD phase transition.

\section{QCD sum rule result}

The QCD sum rule for quarkonium was found to be very reliable at zero
temperature. This is because the expansion parameter of the OPE for the
correlation function for the heavy quark current-current correlation
function is $\Lambda_{\rm QCD}^2/(Q^2+4m_c^2)$.   Therefore, the OPE for the
correlation
function can be reliably estimated  even at $Q^2=0$.  Moreover, for the
heavy quark system, only gluon operators appear, for which reliable
estimates can be made at least for the lowest dimension 4 operator.

How to generalize the sum rule method to finite temperature  depends on
the magnitude of the temperature.  If the temperature is larger than the
separation scale $Q^2+4m_c^2$, we have to calculate the temperature
effect into the Wilson coefficient.  If the temperature is low, then all
the temperature effects can be put into the temperature dependent
operators.  In this case, the new expansion parameter in the OPE will be
$(\Lambda_{\rm QCD}+cT)^2/(Q^2+4m_c^2)$, where $c$ is some constant.
Whether such approximation is valid or not can be directly checked by looking
at the convergence of the OPE at finite temperature.  An additional
ingredient at finite temperature is that unlike at zero temperature,
where only scalar gluon operators appear, operators with Lorentz indexes
have to be added.  Therefore, up to dimension 4,
both the scalar gluon $(G_0)$ and twist-2 gluon $(G_2)$ operators
contribute.  Since we have extracted the temperature dependence of these
operators from the lattice, no ambiguities exist for the OPE side of the
sum rule for a consistent analysis.

\begin{figure}[tb]
\begin{center}
\includegraphics[width=0.5\textwidth]{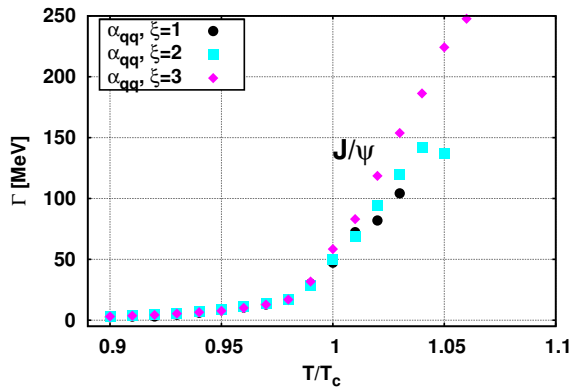}
\end{center}
\caption{Thermal width of $J/\psi$ obtained from the second-order Stark effect and QCD sum rule constraint.
}
\label{thermal-width}
\end{figure}

The OPE for the correlation function of heavy quark currents is
related to the phenomenological side via the dispersion relation.  In
the sum rule, one has to assume the form of the spectral density.  Here,
one notes that lattice calculations suggest that the peak structure in
the spectral density persists above $T_{\rm c}$, although the resolution is
not good enough for a critical study.   Therefore, we can assume a
relativistic Breit-Wigner form for the spectral density.  Using the
moments for the correlation function, we can extract a constraint for
the temperature dependence for the mass and width from the temperature
dependence of the condensates.  We find that if there is no change in
the width, the mass critically decreases by few hundreds MeV slightly
above $T_c$, equivalently if there is no change in mass, the width will
increase by a similar amount~\cite{morita_jpsi}.  The dots in Fig. 2 represent the
maximum mass shift assuming no change in the width.  This is a direct
consequence of the critical change of the scalar condensate proportional
to $\varepsilon-3p$.  As can be seen in the figure, the mass
shift obtained from the second-order Stark effect is almost the same as
the maximum mass shift obtained in the sum rule up to $T_{\rm c}$ and then
becomes smaller. The mass shift at $T_{\rm c}$ is about $-50$ MeV.  In the QCD
sum rules, only a constraint for the combined mass shift and thermal
width could be obtained. This constraint can be crudely summarized as follows,
\begin{eqnarray}
-\Delta m+\Gamma_T \simeq 80+17\times(T-T_{{\rm c}}) \quad {\rm [MeV]},
\label{constraint}
\end{eqnarray}
within the temperature range from $T_{{\rm c}}$  to $1.05T_{{\rm c}}$.
Therefore, the difference between the Stark effect and the maximum mass
shift obtained from QCD sum rules above $T_{{\rm c}}$ in
Fig.~\ref{mass-shift} could be
attributed to the non-perturbative thermal width at finite temperature.
In Fig.~\ref{thermal-width}, we plot the thermal width obtained from
combining the constraint in \eref{constraint} with the mass shift
obtained from the QCD second-order Stark effect. As can be seen in the
figure, the thermal width at 1.05$T_{\rm c}$ becomes larger than 100 MeV.
Such width slightly above $T_{\rm c}$ is larger than that estimated from a
perturbative LO and NLO QCD matrix element calculation together with an
assumption of weakly interacting quarks and gluons with thermal
masses~\cite{Park,song08}, but smaller than a recent phenomenological
estimate~\cite{mocsy07}.

The mass of quarkonium at finite temperature
was also investigated in the potential models~\cite{alberico07}, where
the mass was found to decrease at high temperature.  However, the
potential has to be extracted from the lattice at each temperature and
hence a more detailed investigation are needed to identify the critical
behaviour of $J/\psi$ in the temperature region near $T_{\rm c}$.

%%%%%%%%%%%%%%%%%%%%%%%%%%%%%%%%%%%%%%%
%%%%%%%%%%%%%%%%%%%%%%%%%%%%%%%%%%%%%%%

\section{Corrections from dynamical quarks}

To consider the quark effects, first we consider the quark operators
appearing in the OPE. We can neglect the light quark contribution
to the OPE,  because the light quark operators appear in the OPE at
order $\alpha_s^2(q^2)$: This is why the light quark condensate can be
neglected in the sum rules for heavy quark system.  On the other hand,
thermal heavy quarks that directly couple to the heavy quark current
contribute to the OPE at leading order.  This is different from the
heavy quark condensates that are perturbatively generated from the gluon
condensates, and contribute to the OPE through gluon condensates, whose
Wilson coefficients are calculated in the momentum representation.  The
direct thermal quark contributions are called the scattering terms.
However, similar terms also appear in the phenomenological spectral
density, which also has free charm quark mode that is not coupled with a
light quark in the form of a $D$ meson above $T_{{\rm c}}$ as been
recently studied in Ref.~\cite{Dominguez}.  Therefore, the scattering
term will cancel out between the OPE and the phenomenological spectral
density in the QCD sum rule analysis for the deconfined medium.

\begin{figure}[ht]
 \begin{center}
 \includegraphics[width=0.5\textwidth]{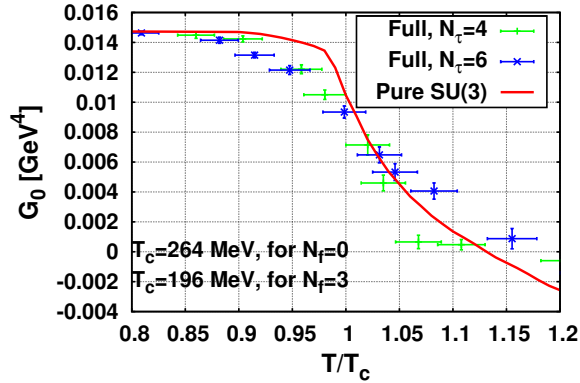}
 \end{center}
 \caption{Comparison of the scalar
 gluon condensate in the pure gauge theory with the one extracted from full lattice QCD.
 Horizontal errorbars in the full QCD case are drawn by assuming the 2\%
 uncertainty in the conversion from the lattice units to physical temperature
 \cite{cheng08}.}
 \label{fig:e3p_full}
\end{figure}

Second, the gluon condensates itself can have a different temperature
dependence in the presence of dynamical quarks.  As discussed before,
the important input for the mass and width change is the temperature
dependence of gluon condensates in Fig.~\ref{fig:gc}; in particular the
dominant contribution comes from the temperature dependence of $G_0$.
For that purpose, we note that the trace of the energy momentum tensor
to leading order is given as,
\begin{equation}
 T^{\mu}_\mu = - \left( \frac{11-2/3 N_{\rm f}}{8} \right)
  \left\langle \frac{\alpha_{\rm s}}{\pi}
   G^{a}_{\mu\nu}G^{a\mu\nu} \right\rangle
    + \sum_q m_q \langle \bar{q}q \rangle. \label{eq:total_trace}
\end{equation}
Therefore, we start from the lattice calculation of the trace of the energy momentum tensor for the full QCD with realistic quark masses as given in Ref.~\cite{cheng08}.   Then, we subtract the
fermionic part of the trace anomaly, which was also shown in the
literature, from the total.  Then we divide the result for the relevant
prefactor with $N_{\rm f}=3$ multiplying the gluon condensate as given in
\eref{eq:total_trace}.   Since the critical
temperature $T_{{\rm c}}$ differs, we compared it as a function of
$T/T_{\rm c}$ in
which $T_{{\rm c}}=196$ MeV for the full QCD case~\cite{cheng08}.
As can be seen in  Fig.~\ref{fig:e3p_full}, the magnitude of the
resulting change near the critical temperature are remarkably similar
between the full and pure gluon QCD; although the slope at $T_{{\rm c}}$ is
milder for full QCD as a consequence of rapid cross over transition
instead of a first order phase transition.
 Since the change of the condensate sets in at a lower $T/T_{{\rm c}}$
 in the full QCD case,  the mass and width of charmonia might start
 varying at a lower temperature in the realistic case than in the pure
 glue theory.  Therefore we
believe our main argument and the quantitative result will not be alter
even in the realistic situation.

\section{Summary}

We have shown that the critical mass shift  and width increase of quarkonia slightly above $T_c$ could effectively be an "order parameter" of QCD phase transition.   The expected shift for $J/\psi$ is small and will be a challenge for LHC.  However, larger shift is expected for $\chi_c$ at its formation temperature slightly above $T_c$.  Therefore, direct measurements and confirmation is possible.  The changes will also lead to changes in production ratios in the statistical model and $J/\psi$ suppression effects, which needs further detailed studies.  This work was supported by BK21 Program of the Korean Ministry of
 Education, and by the Korean Research Foundation
 KRF-2006-C00011.

\section*{References}
\end{document}